\documentclass[twocolumn,showpacs,floatfix,aps]{revtex4}
\usepackage[dvips]{graphics}

\def \beq {\begin{equation}}
\def \eeq {\end{equation}}

\usepackage{amssymb}

\begin{document}

\title{Lifetime and decay of unstable particles  in strong gravitational fields}

\author{Douglas Fregolente}
\email{douglasf@ime.unicamp.br}
\affiliation{
Departamento de Matem\'atica Aplicada,
IMECC -- UNICAMP, \\
C.P. 6065, 13083-859 Campinas, SP, Brazil.}
\author{Alberto Saa\footnote{On leave of absence from UNICAMP, Campinas, SP, Brazil.}}
\email{asaa@ime.unicamp.br}
\affiliation{Centro de Matem\'atica, Computa\c c\~ao e Cogni\c c\~ao,
Universidade Federal do ABC, 09210-170  Santo Andr\'e, SP, Brazil}
\pacs{95.30.Cq, 14.20.Dh}

\begin{abstract}
We consider here the decay of unstable particles in geodesic circular motion
around compact objects.
For the neutron, in particular,
strong and weak decay
 are calculated by means of
a semiclassical approach.
Noticeable effects are expected to occur as one approaches the
photonic circular orbit of realistic black-holes. We argue that,
in such a limit,
  the quasi-thermal spectrum inherent to
extremely relativistic observers in circular motion plays a   role similar to the
Unruh radiation for uniformly accelerated observers.
\end{abstract}

\maketitle

\section{Introduction}

The possible decay of inertially stable particles due to strong gravitational
fields has been considered recently. In particular, the proton decay, by weak and strong interactions,
in uniformly accelerated trajectories\cite{unif1,unif2,uniform} and
 in circular
motion around compact objects\cite{circ},
has been considered in great detail. The astrophysical implications of these
results are now under investigation. The possible decay of accelerated protons, however, is
not a new issue. It can be traced back to the works of Ginzburg and Zharkov\cite{GZ} in the
sixties, where processes of the type
$
p^+ \stackrel{a}{\longrightarrow} n^0\pi^+
$
were considered. At the same time, Zharkov\cite{Z} investigated the weak and strong
decay of protons accelerated by
an external electromagnetic field. (See \cite{R} for
a review.)
Clearly, none of these processes would be   allowed in the absence of external forces.
We notice, however, that
 there are   subtle differences  between processes involving uncharged particles
 where the accelerations
 have gravitational and electromagnetic origins, see \cite{circ} for further details.

In this paper, we consider the decay and the lifetime of
unstable particles in geodesic circular motion
around spherically symmetrical compact objects.
We evaluate, in particular,
 decay rates and lifetime   for neutrons
 in relativistic circular motion according to
the  semiclassical approach introduced in \cite{circ}.
Both the weak
\beq
\label{weak}
n^0\stackrel{a}{\longrightarrow} p^+e^-\bar{\nu}_e
\eeq
and the (inertially forbidden) strong
\beq
\label{strong}
n^0\stackrel{a}{\longrightarrow} p^+\pi^-
\eeq
channels are considered.
Our results are compared to several ones obtained previously in the literature for the
related processes
$
p^+  \stackrel{a}{\longrightarrow}  n^0e^+{\nu}_e $ and
$p^+  \stackrel{a}{\longrightarrow}  n^0\pi^+ $.
As we will see, for realistic black-holes,
noticeable effects are expected to occur
for circular geodesics close to the
photonic  orbit $r=3GM/rc^2$. Observers in these (unstable) circular orbits are necessarily
in extremely relativistic motion $(v^2\approx c^2)$, and it is well known
that they indeed realize the inertial vacuum as a quasi-thermal distribution
of particles characterized by a temperature $T$ in the range\cite{temp,review}
\beq
\label{temp}
\frac{\hbar a}{4\sqrt{3}c} \le kT \le \frac{\hbar a}{2\sqrt{3}c},
\eeq
where $a$ stands for the effective Minkowskian centripetal acceleration for
 relativistic circular orbits.
Our results suggest that
the temperature (\ref{temp}) have in the present case
 the same central role played
by  Unruh temperature\cite{unruheffect}
in the analysis of   uniformly accelerated
 particles as seen from Rindler observers\cite{uniform}. (See \cite{review} for a recent review.)
 Similar conclusions hold also for other unstable particles.

\section{Circular geodesics around compact objects}

The  line element corresponding to  a spherically symmetrical object  of mass $M$ is given by the Schwarzschild metric
\beq
\label{schwarz}
ds^2 = -\left(1- {2M}/{r} \right)dt^2 + \left(1- {2M}/{r} \right)^{-1}dr^2 + r^2d\Sigma^2,
\eeq
where $d\Sigma^2 = d\theta^2 + \sin\theta d\phi^2$. Natural unities are adopted hereafter.
In this coordinate system\cite{wald}, a particle of mass $m$ in a circular timelike geodesic at radius
$r$ on the equatorial plane   have energy per mass ratio given by
\beq
{\cal E}/m = (1-2M/r)\dot{t} = (1-2M/r)/\sqrt{1-3M/r},
\eeq
with the dot standing for $s$-derivative.
Its angular momentum $L=r^2\dot{\phi}$ can be calculated directly  from the
definition of a timelike circular  geodesic  parameterized by the proper time $s$,
%\beq
%-1 = - \left(1-2M/r \right)\dot{t} + r^2\dot{\phi}^2,
%\eeq
leading finally to the following expression for the worldline of a timelike circular
equatorial geodesic
   in
Schwarzschild coordinates
\beq
\label{circ}
x^a(s) = \left(\frac{s}{\sqrt{1-3M/r}},r,\pi/2,s\sqrt{\frac{ {M/r^3}}{1-3M/r} }  \right).
\eeq
Clearly, since the trajectory (\ref{circ}) is a geodesic, its acceleration $a^b = \dot{x}^c\nabla_c \dot{x}^b$
calculated with respect to the metric (\ref{schwarz})
vanishes identically. However, we will proceed here
 in a different manner.
In the next section, we will consider quantum effects as realized by observers
with circular trajectories as
(\ref{circ}) in the Minkowski spacetime.
An observer following the worldline   (\ref{circ}) in Minkowski spacetime
experiences a centripetal acceleration
\beq
\label{accel}
a = \sqrt{a_ba^b} = \frac{M/r^2}{1-3M/r}.
\eeq
On the other hand,
in Minkowski spacetime   the worldline
of a particle in  a uniform circular motion on the equatorial plane
with angular velocity $\Omega$  is given by
\beq
\label{trajec}
x^a(\tilde{s}) =  \left(t,r,\pi/2,\Omega t   \right),
\eeq
from which one has immediately $\dot{x}^a=\gamma(1,0,0,\Omega)$ and
$a=\sqrt{a_ba^b} = r\gamma^2\Omega^2$, where the constant $\gamma = dt/d\tilde{s}=(1-r^2\Omega^2)^{-1/2}$
corresponds to the Lorentz factor. The angular velocity $\Omega$ is to be determined by imposing
the centripetal acceleration (\ref{accel}) for the trajectory (\ref{trajec}), yielding
\beq
\label{pseudoaccel}
\Omega =   \sqrt{\frac{M/r^3}{1-2M/r}}
\eeq
and the following Lorentz factor
\beq
\label{pseudoaccel1}
 \gamma =   \sqrt{\frac{1-2M/r}{1-3M/r}}.
\eeq

The treatment of the quantum effects   realized by observers in the circular geodesic (\ref{circ})
of Schwarzchild spacetime by means of an effective Minkowskian circular trajectory is, of
course, only an approximation. It is shown in \cite{comp}, nevertheless,
 that the results obtained in a semiclassical approach assuming a Schwarzschild
 spacetime
and a flat spacetime with external ``Newtonian'' attraction forces   such that (\ref{accel}) and
(\ref{pseudoaccel})
 hold  differ by no more than
30\%, if we restrict ourselves to the circular orbits with $r>3M$. Since Schwarzschild
spacetime is asymptotically flat, it
is indeed natural that the emitted powers calculated  in Minkowski and
Schwarzschild
spacetime agree    when one considers
circular motions with large $r$. In fact, as it is shown in \cite{circ} and \cite{comp},
the power emitted by a particle in circular motion with radius $r$  in Minkowski spacetime
with angular velocity (\ref{pseudoaccel})
is very close to that one emitted by a particle in
a circular geodesic with the same radius $r$ in a Schwarzschild
spacetime, provided that $r>6M$. This is in agreement with the well known
fact that processes involving wavelengths with the same  order of magnitude of the
  Schwarzschild
radius   need necessarily to be analyzed  using fully curved spacetime
calculations. Moreover, the acceleration defined in (\ref{accel}) has the additional
desirable feature of being divergent at $r=3M$, in accordance with previous
works on geodesic emission\cite{Misneretal}, which established that near the
photonic orbit the emitted power diverges.
The angular velocity (\ref{pseudoaccel}) mimics
the main qualitative properties of the real Schwarzschild circular geodesics,
justifying our assumption of circular trajectories in
a flat spacetime with centripetal acceleration   (\ref{accel}).

\section{Emission rates and lifetimes}

The semiclassical current formalism employed in \cite{circ} to the proton decay case
consists basically in considering the proton and the neutron as
distinct energy eigenstates $|p\rangle$ and $|n\rangle$ of a two-level system
such that $\hat{H}_0 |p\rangle = m_p |p\rangle $ and $\hat{H}_0 |n\rangle  = m_n |n\rangle$,
%\beq
%\label{2levels}
%\hat{H}_0 |p\rangle = m_p |p\rangle ,\quad \hat{H}_0 |n\rangle  = m_n |n\rangle,
%\eeq
where $\hat{H}_0$ is the proper Hamiltonian of the system, $m_p$ and $m_n$  are, respectively,
the proton and neutron masses. The weak channel (\ref{weak}) is implemented by considering a vector current
associated to the two-level system  coupled to a   quantized fermionic field
(corresponding to the electron $e^-$ and to the anti-neutrino $\bar{\nu}_e$) by means
of the effective coupling constant $G_{\rm w}$, which is about the
 order of the
Fermi coupling constant $G_F \approx 1.166\times 10^{-5} {\rm\, GeV}^{-2}$, whereas
the strong  channel (\ref{strong}) involves a scalar current
coupled to a   quantized bosonic field (the pion $\pi^-$)
by means
of the effective coupling constant $G_{\rm s}$, of the  order of the  pion-nucleon-nucleon strong coupling $g_{\pi \!N\!N}^2/4\pi \approx 14$\cite{nuclear}.
The currents are then specialized to the case of uniform circular trajectories with radius
$r$ and angular velocity $\Omega$ (and centripetal acceleration $a=r\gamma^2\Omega^2$)
in Minkowski spacetime.

\begin{figure}[ht]
\resizebox{1\linewidth}{!}{\includegraphics*{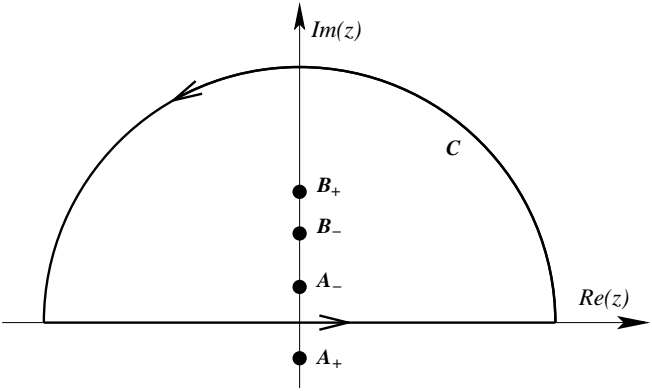}}
\caption{Path   used in the complex integrations (\ref{decay1}) and (\ref{decay2}).
Both terms (\ref{protonrate1}) and (\ref{protonrate2}) come from the pole $B_+$. They coincide
with the values calculated in \cite{circ} by integrating around the pole $A_+$.
Notice that, for the processes considered there,   the term
$\delta$ has a different sign.
 The second terms in
(\ref{decayw}) and (\ref{decays}) come  from the (degenerated) poles $A_-$ and
$B_-$. }
\label{fig1}
\end{figure}

The proper decay rates  corresponding to the weak   (\ref{weak}) and to
the strong (\ref{strong}) channels are given, respectively, by
\begin{eqnarray}
\label{decay1}
\Gamma_{n\rightarrow p}^{\rm w}  &=&  - \frac{G^2a^5}{8\pi^4 }\oint d\lambda \,
e^{i\delta }
 \frac{A_{(bc)}Z^bZ^c}{(Z_aZ^a)^2} \times \nonumber \\ 
 && \quad\quad  \left(\frac{16}{\gamma^4(Z_aZ^a)^2} + 4 \frac{\mu^2}{\gamma^2 Z_aZ^a } \right)
\end{eqnarray}
and
\beq
\label{decay2}
\Gamma_{n\rightarrow p}^{\rm s} = - \frac{G^2a}{4\pi^2}\oint d\lambda \frac{e^{i\delta }}{\gamma^2Z_aZ^a},
\eeq
where  $\mu^2 = (m_e^2+m_\nu^2)/a^2$, $m_e$ and $m_\nu$ being, respectively,
 the electron and neutrino masses, and $\delta = (m_n - m_p)/a$. We assume here $m_\nu=0$.
The rates (\ref{decay1}) and (\ref{decay2}) are
obtained, respectively, from equations (3.31) and (3.16) of \cite{circ},
where all the relevant details
can   be found.
In particular, we have $Z^a=(-\lambda + i\epsilon, 0, -(2ra/\gamma)\sin(\Omega\lambda\gamma/2a),0)$,
where $0<\epsilon \ll 1$ is a regulator.
For the relativistic case ($\gamma \gg 1$), the terms involved in the integrations
(\ref{decay1}) and (\ref{decay2}) are\cite{circ}
\begin{eqnarray}
\label{apro1}
A_{(bc)}Z^bZ^c  &\approx&  \frac{\lambda^2}{\gamma^4 }\left(\frac{\lambda^4}{72} + \frac{\lambda^2}{12} + 1 \right),\\
 Z_aZ^a &\approx& \frac{1}{12\gamma^2}
\left( \lambda + i\sqrt{3}A_+\right)
\left( \lambda + i\sqrt{3}A_-\right)\times \nonumber \\
&&\quad 
\left( \lambda - i\sqrt{3}B_+\right)
\left( \lambda - i\sqrt{3}B_+\right),
\label{apro2}
\end{eqnarray}
where
$A_\pm = 1\pm \sqrt{1+2\epsilon/\sqrt{3}}$,
$B_\pm = 1\pm \sqrt{1-2\epsilon/\sqrt{3}}$, see Fig. \ref{fig1}.

In contrast to the case considered in \cite{circ}, since $m_n > m_p$,
one needs here to perform the complex integrations (\ref{decay1}) and
(\ref{decay2}) along the path depicted in
Fig. \ref{fig1}.
 We get, after taking the limit    $\epsilon\rightarrow 0$,
 \beq
 \label{decayw}
 \Gamma_{n\rightarrow p}^{\rm w} =
 \Gamma_{p\rightarrow n}^{\rm w} +  \frac{G_{\rm w}^2a^5\delta}{90\pi^3}  \left(20 + 15 \delta^2 + 3  \delta^4
 -15\mu^2\left( 1+  \delta^2 \right) \right)
 \eeq
 and
 \beq
 \label{decays}
\Gamma_{n\rightarrow p}^{\rm s} =  \Gamma_{p\rightarrow n}^{\rm s} + \frac{G_{\rm s}^2a\delta}{2\pi},
\eeq
where $\Gamma_{p\rightarrow n}^{\rm w}$ and $\Gamma_{p\rightarrow n}^{\rm s}$ correspond,
respectively, to
the proper decay rates associated to the inverse processes   considered in \cite{circ},
\begin{eqnarray}
\label{protonrate1}
\Gamma_{p\rightarrow n}^{\rm w} &=& \frac{G_{\rm w}^2a^5 e^{-2\sqrt{3}\delta }}{1728\pi^3} \left[
 49\sqrt{3} + 102 \delta  +30\sqrt{3}\delta^2
 +12 \delta^3 \right.
      \nonumber \\
   & & \quad\quad \left.  - \mu^2\left(39\sqrt{3}   + 90\delta
 +36\sqrt{3}\delta^2  \right) \right]
\end{eqnarray}
and
\beq
\label{protonrate2}
\Gamma_{p\rightarrow n}^{\rm s} =  {G_{\rm s}^2ae^{-2\sqrt{3}\delta }}/({8\sqrt{3}\pi}).
\eeq
The approximations involved in the derivation of the expressions (\ref{decay1}) and (\ref{decay2})
require, respectively, that the centripetal acceleration $a$ obeys $m_e<a<m_p$ and
$m_\pi <a< m_p$, where $m_\pi$ stands for the pion $\pi^-$ mass.
We notice that, for accelerations $a\gg m_p$,  the no-recoil hypothesis is violated\cite{circ} and,
hence, our approximation  breaks down.
The neutron proper lifetime associated with the decay rates (\ref{decayw}) and (\ref{decays}) are
given simply by $\tau^{\rm w}_n = 1/\Gamma^{\rm w}_{n\rightarrow p}$ and
$\tau^{\rm s}_n = 1/\Gamma^{\rm s}_{n\rightarrow p}$. Our results assume a particularly simple form
if one considers the proton and neutron lifetime ratio, namely
\beq
\label{sratio}
 {\tau_p^{\rm s}}/{\tau_n^{\rm s}} = 1 + 12 \delta  e^{2\sqrt{3} \delta },
 %\, {\rm and} \,
 %{\tau_p^{\rm w}}/{\tau_n^{\rm w}} =  1 + \frac{96}{5}\delta e^{2\sqrt{3} \delta } %\frac{P(\delta,\mu)}{Q(\delta,\mu)}  ,
\eeq
and
\beq
\label{wratio}
 {\tau_p^{\rm w}}/{\tau_n^{\rm w}} =  1 + \frac{96}{5}\delta e^{2\sqrt{3} \delta } \frac{P(\delta,\mu)}{Q(\delta,\mu)}  ,
\eeq
where  $P(\delta,\mu )$
%\beq
% P(\delta,\mu ) =   20 + 15 \delta^2 + 3  \delta^4
% -15\mu^2\left( 1+  \delta^2 \right)
%\eeq
and $Q(\delta,\mu )$
%\begin{eqnarray}
% Q(\delta,\mu ) &=&  49\sqrt{3} + 102 \delta  +30\sqrt{3}  \delta^2
% +12 \delta^3 \nonumber \\ && -\mu^2\left(39\sqrt{3}   + 90 \delta
% +36\sqrt{3} \delta^2  \right)
%\end{eqnarray}
are polynomials easily obtained from (\ref{decayw}) and (\ref{protonrate1}).
It is clear that for large $a$, both ratios obey
\beq
\label{asymp}
\tau_p/\tau_n  \approx 1 + O(a^{-1}).
\eeq
The strong
channel is expected to be the dominant decay mode for neutrons with centripetal
 accelerations $a$ such that
$m_\pi  <a< m_p$. Taking into account that $m_\pi \approx 139.57$ MeV, $m_n \approx 939.56$ MeV and that
$m_p \approx 938.27$ MeV, we have that the proton and the neutron lifetimes differ by no more than
1\% to 10\%
in the range of accelerations where the strong channel dominates.
The weak channel, on the other hand, dominates for smaller accelerations
 $m_e<a< m_\pi$. (We remind that $m_e\approx 0.51$
MeV.) From  (\ref{sratio}) and (\ref{wratio}), it is clear that for
\beq
\label{critical}
a \gg a_c =  m_n - m_p   \approx 1.29 {\rm\, Mev}
\eeq
the asymptotic expression (\ref{asymp}) holds accurately.
The meaning of the ``critical'' acceleration $a_c$ will
be discussed in the last section. Here, we mention only that $a_c$ belongs to the range where
the weak channel dominates. In fact, for $m_e < a < a_c$, a significative difference between the
proton and the neutron lifetime is observed.

\begin{figure}[ht]
\resizebox{1\linewidth}{!}{\includegraphics*{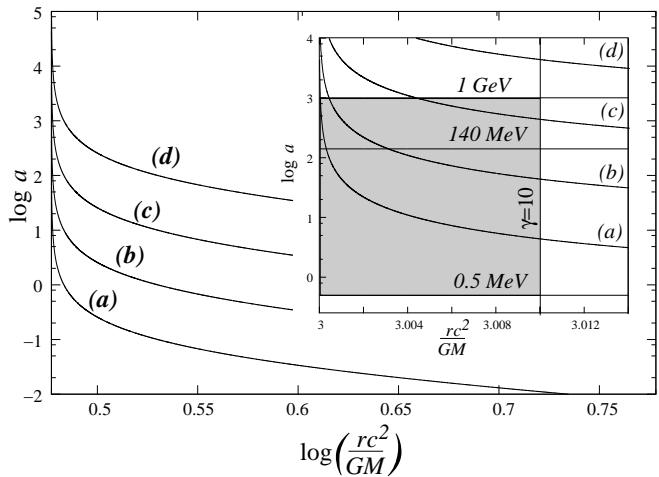}}
\caption{
Centripetal acceleration, in MeV, given by   equation (\ref{accel1}),
  for circular trajectories such that
$3GM/c^2 < r \le 6GM/c^2$. The curves (a) to (d) correspond, respectively, to the ratios
$M/M_\odot = 10^{-15}, 10^{-16}, 10^{-17}, $ and $10^{-18}.$ We recall that, for the
neutron decay, the approximations involved in the derivation of
our results require $a < m_p \approx 938.27$ MeV  (no-recoil hypothesis) and
$a > m_e \approx 0.51$ MeV (weak channel) or
$a > m_\pi \approx 139.57$ MeV (strong channel). The
hypothesis of an extremely relativistic motion ($\gamma \gg 1$), on the
other hand, requires orbits close to the photonic one, see equation (\ref{gamma}).
In the detail, the hatched area denotes the region of validity of our approximations,
assuming $\gamma>10$.
 }
\label{figaccel}
\end{figure}

A proper acceleration of the order of 1 MeV is extremely high. For sake of comparison, protons in
the CERN Large Hadron Collider have proper acceleration $a\approx 10^{-8}$ MeV\cite{circ}.
In order to compare $a_c$ with centripetal accelerations induced by realistic black-holes, we   cast
(\ref{accel}) in the form
\beq
\label{accel1}
a \approx 1.34\times 10^{-16}
\left(\frac{M_{\odot}}{M}\right)\frac{\left(GM/rc^2\right)^2}{1-3GM/rc^2 },
\eeq
 where $a$ is now measured
in MeV and the numerical constant corresponds to
 $\hbar c^3/GM_\odot$.
 Hence, realistic black holes $(M\approx M_\odot)$ will induce centripetal accelerations
of the MeV order only for those circular orbits very close to the photonic orbit $r=3GM/c^2$.
Smaller black holes, nevertheless, can induce considerably higher centripetal accelerations for
the (unstable) circular orbits located between the photonic orbit and the last stable circular
orbit at $r=6GM/c^2$, see Figure \ref{figaccel}.
On the other hand, the hypothesis of an extremely relativistic motion
($\gamma \gg 1$) adopted in the approximations (\ref{apro1}) and (\ref{apro2})
requires circular trajectories close to the photonic orbit, since from (\ref{pseudoaccel1}) one
has
\beq
\label{gamma}
\frac{rc^2}{GM} = \frac{3\gamma^2-2}{\gamma^2-1}.
\eeq

\begin{figure}[ht]
\resizebox{1\linewidth}{!}{\includegraphics*{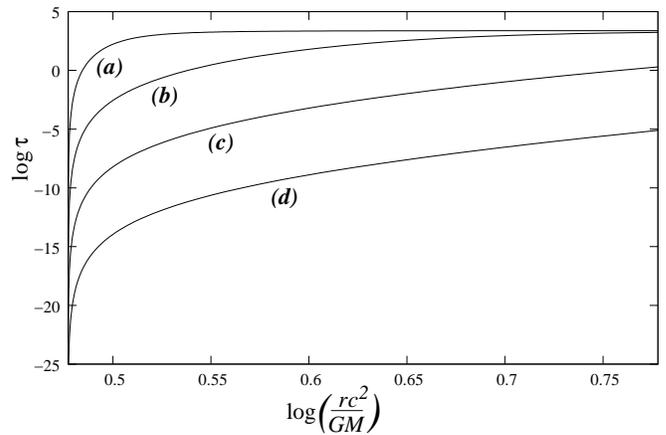}}
\caption{Neutron proper lifetime $\tau$, in seconds, for circular trajectories such that
$3GM/c^2 < r \le 6GM/c^2$. The curves (a) to (d) correspond, respectively, to the ratios
$M/M_\odot = 10^{-16}, 10^{-17}, 10^{-18}, $ and $10^{-19}.$ The (free) neutron inertial lifetime is
approximately $886 \, $s. The   validity region of our approximations is depicted in
Fig. 2.}
\label{fig2}
\end{figure}

\begin{figure}[ht]
\resizebox{1\linewidth}{!}{\includegraphics*{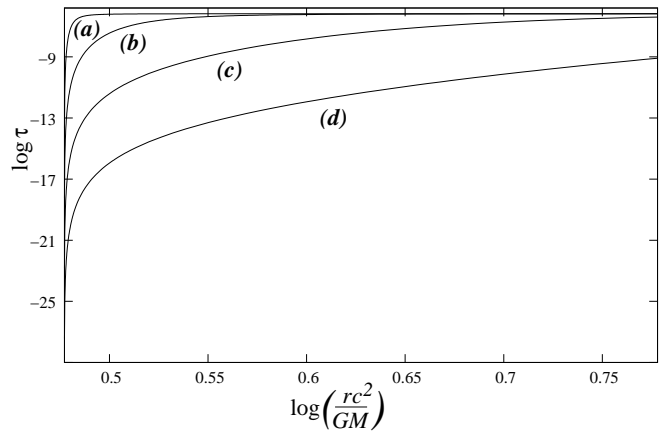}}
\caption{Muon proper lifetime $\tau$, in seconds, for circular trajectories such that
$3GM/c^2 < r \le 6GM/c^2$. The curves (a) to (d) correspond, respectively, to the ratios
$M/M_\odot = 10^{-17}, 10^{-18}, 10^{-19}, $ and $10^{-20}.$ The muon inertial lifetime is
about $2.2\times 10^{-6}$s  and
 the branching ratio corresponding to the inertial process $\mu^- \rightarrow e^-\bar{\nu}_e\nu_\mu$
is greater than 98\%. The  validity region of our approximations is depicted in
Fig. 2.
}
\label{fig3}
\end{figure}

 Figure \ref{fig2}
depicts the lifetime for a neutron due to weak decay
in   circular orbits with $3GM/c^2 < r \le 6GM/c^2$ around
small black holes.
As expected, the smaller is the black hole, the larger is the reduction in the particle lifetime.
The semiclassical approach used here can be applied for other unstable particles as well.
Particularly interesting is the muon weak decay
$
\mu^- \stackrel{a}{\longrightarrow} e^-\bar{\nu}_e\nu_\mu,
$
which can also be described by a vector current based on a two-level system  coupled to
quantized fermions (the neutrinos $\bar{\nu}_e$ and $\nu_\mu$) by means of a coupling
constant of the same order than $G_F$. Since neutrinos have very small masses, the
rate (\ref{decayw}) for the muon weak decay is accurate for accelerations in the
range $0<a<m_e \approx 0.51$ MeV. (We remind that  $m_\mu \approx 105.7$ MeV.)
Figure \ref{fig3}
depicts
the lifetime of a muon
in  geodesic circular orbits of
small black holes such that $3GM/c^2 < r \le 6GM/c^2$

\section{Discussion}

As Figures \ref{fig2} and \ref{fig3} show,   small black holes are necessary in order  to induce
sensitive alterations in the proper lifetime of unstable particles in circular
orbits. However, noticeable effects do occur in the vicinity of the photonic orbit
for realistic black hole. Despite that the analysis presented here is restricted to the
(unstable) circular geodesics close to $r=3M$, it can give some hints about
the behavior of particles in more realistic situations.
As equations (\ref{sratio}) and (\ref{wratio})  reveal, free protons and neutrons
in geodesic circular motion close to the photonic orbits have comparable proper lifetime.
This situation is completely different from the inertial one, and its implication to
particle physics in the vicinity of black holes has not been sufficiently studied yet.
A similar conclusion holds for the muon. For a  accelerations $a\gg a_c=2\sqrt{3}(m_\mu - m_e)= 364.4$ MeV,
neglecting possible effects of back-reaction\cite{circ},
the muon and the electron, which in such a case can indeed decay by the inverse process $
e^- \stackrel{a}{\longrightarrow} \mu^-\bar{\nu}_\mu\nu_e,
$
have comparable lifetimes.

In order to grasp the meaning of the temperature (\ref{temp}), let us consider the case of
the weak decay of protons and neutron in uniformly accelerated motion, where the Unruh temperature\cite{unruheffect}
$T_U =  a/2\pi$ is known to play a central role\cite{uniform,review}.
The proton and neutron lifetime ratio for this
  case can be obtained from the decay rates (3.13) and (3.17) of reference \cite{unif2},
\beq
\label{ee}
 {\tau_p^{\rm w}}/{\tau_n^{\rm w}} = e^{2\pi\delta},
\eeq
which also has the asymptotic form (\ref{asymp}) for large values of $a$. Notice that,
in this case,
we have exactly  the same
critical acceleration $a_c$ of (\ref{critical}).

In the
case of the uniformly accelerated motion, one can   describe the decay of protons and neutrons
as seen by comoving Rindler observers. The key point here is that Rindler observers realize the
inertial vacuum as a thermal state with temperature $T_U =  a/2\pi$.
Heuristically,
one can imagine the two-level system in thermal equilibrium with the Unruh radiation
associated with the quantized fields in question.
For
our two-level system    in thermal equilibrium at temperature $T$, the probability
of occupation of the proton $|p\rangle$   and neutron   $|n\rangle$ states are, respectively
\beq
N_p = \frac{e^{-m_p/T}}{e^{-m_p/T}+e^{-m_n/T}},\quad
N_n = \frac{e^{-m_n/T}}{e^{-m_p/T}+e^{-m_n/T}}.
\eeq
The ratio $N_p/N_n = e^{(m_n-m_p)/T}$ diverges for $T\rightarrow 0$, indicating that for
low temperatures, the system is likely to be in its fundamental state. However, for temperatures
$T\gg(m_n-m_p) $ the ratio  tends to 1, indicating that the system can be in the states
$|p\rangle$ or $|n\rangle$  with equal probability. In other words, the transitions
$|p\rangle\rightarrow |n\rangle $ and $|n\rangle\rightarrow |p\rangle$ become equally probable for high temperatures, shedding some
light in the expression (\ref{ee}). For linear accelerations $a$ such that the associate Unruh
temperature $T_U =  a/2\pi$ is much higher than the energy gap $m_n-m_p$, it is natural
to expect that protons and neutrons have the same lifetime, since both transitions
of the two level systems  are equally probable. The lifetime ratios (\ref{sratio}) and
(\ref{wratio})
 suggest something similar to the case of uniform circular trajectories.
They can be understood if one considers the two-level system in equilibrium with the
quasi-thermal radiation with temperature $(\ref{temp})$ associated with the quantized
fields in question, confirming
the view that observers in relativistic circular motion
with centripetal acceleration
$a$ do  realize the inertial vacuum as a quasi-thermal state with temperature (\ref{temp})
for the extremely relativistic case\cite{temp,review}.
We stress that this state is quasi-thermal in the sense that it can be described
by a temperature that varies slowly with the energy gap $\Delta E$ of the
two-level system, monotonically
from $(\pi/2\sqrt{3})T_U$ to $(\pi/\sqrt{3})T_U$, corresponding respectively
  to low and to high values of $\Delta E/a$ \cite{unruh,review,milgrom}.
  For the neutron decay, for instance,
  $\Delta E = m_n-m_p \approx 1.29$ MeV, implying that for circular
  trajectories close to the photonic orbit $r = 3GM/c^2$, the quasi-thermal state
  is indeed characterized by a temperature close  to the lower bound of (\ref{temp}).
  For the muon decay, on the other hand, the temperature is closer to the upper
  bound of (\ref{temp}).
  A definitive answer to this problem, however, must necessarily face the subtle issue of
   quantum thermal distributions for rotating systems\cite{ottewill}.

\acknowledgements
The authors are grateful to G. Matsas for enlightening discussions.
This work was supported by FAPESP, CNPq, and UFABC.
A.S.  is grateful  to   Prof. V. Mukhanov
for the warm hospitality at the Ludwig-Maximilians-Universit\"at, Munich, where   part of this work
was carried out.

\end{document}